\RequirePackage{fix-cm}
\documentclass[twocolumn,epjc3]{svjour3}
\smartqed  
\RequirePackage{graphicx}
\journalname{Eur. Phys. J. C}
\usepackage[latin1]{inputenc}
\usepackage{amssymb}
\usepackage{color}
\usepackage{graphicx}
\usepackage{dcolumn}
\usepackage{bm}
\usepackage[header,title,page,titletoc]{appendix}
\usepackage{amsmath}
\usepackage{amsfonts}
\usepackage[colorlinks,citecolor=blue,urlcolor=blue,linkcolor=blue]{hyperref}

\begin{document}

\title{A 331 WIMPy Dark Radiation Model}

\author{Chris Kelso\thanksref{e1,addr1}
       \and
       C. A. de S. Pires\thanksref{e2,addr2}
       \and
       Stefano Profumo\thanksref{e3,addr3}
       \and
       Farinaldo S. Queiroz\thanksref{e4,addr3}
       \and 
       P. S. Rodrigues da Silva\thanksref{e5,addr2}
       }

\thankstext{e1}{e-mail: ckelso@physics.utah.edu}
\thankstext{e2}{e-mail: cpires@fisica.ufpb.br}
\thankstext{e3}{e-mail: stefano@scipp.ucsc.edu}
\thankstext{e4}{e-mail: fdasilva@uscs.edu}
\thankstext{e5}{e-mail: psilva@fisica.ufpb.br}

\institute{Department of Physics and Astronomy, University of Utah, Salt Lake City, UT 84112, USA\label{addr1}
          \and 
          Departamento de F\'{\i}sica, Universidade Federal da Para\'{\i}ba, Caixa Postal 5008, 58051-970, Jo\~ao Pessoa, PB, Brasil \label{addr2}
          \and 
           Department of Physics and Santa Cruz Institute for Particle Physics University of California, Santa Cruz, CA 95064, USA \label{addr3}
          } 
\date{Received: date / Accepted: date}

\maketitle

\begin{abstract} 
Recent observations suggest that the number of relativistic degrees of freedom in the early universe might exceed what is predicted in the standard cosmological model. If even a small, percent-level fraction of dark matter particles are produced relativistically, they could mimic the effect of an extra realistic species at matter-radiation equality while obeying BBN, CMB and Structure Formation bounds. We show that this scenario is quite naturally realized with a weak-scale dark matter particle and a high-scale ``mother'' particle within a well motivated 3-3-1 gauge model, which is particularly interesting for being consistent with electroweak precision measurements, with recent LHC results, and for offering a convincing explanation for the number of generations in the Standard Model. 
\end{abstract}


\section{INTRODUCTION}
\label{introduction}

Dark Matter is a compelling evidence for new physics beyond the Standard Model. Observations from a variety of experiments and physical scales have by now conclusively established the existence of dark matter. Although most of the dark matter must be non-relativistic (or ``cold'') to satisfy observations of how structures form in the universe, in the last few years an intriguing evidence for the existence of additional {\em relativistic} degrees of freedom in the early universe has started to accumulate. Albeit in no sense is the $\Lambda$CDM paradigm being conclusively challenged, the question of how new physics could accommodate this extra relativistic degrees of freedom is by all means intriguing.

The Planck collaboration has recently reported their   accurate measurements of the power spectrum of the cosmic microwave background radiation (CMB) \cite{Planck}. 
The collaboration claimed no evidence for an extra radiation component, which is usually interpreted in terms of the number of relativistic species ($N_{eff}$) at the decoupling of the CMB. 
Indeed, Planck has reported $N_{eff} = 3.36^{+0.68}_{-0.64}$ at $95\%$ C.L together with a fairly low value for the expansion rate of the universe today, $H_0 = (67.3 \pm 1.2)\ { \rm km s^{-1} Mpc^{-1}}$ \cite{Planck}. 
Nevertheless, it  has been pointed out by the Planck collaboration that two recent observations of Cepheid variables by the Hubble Space Telescope (HST) yielded $H_0 = (73.8 \pm 2.4)\ {\rm km s^{-1} Mpc^{-1}}$ which is $2.5\sigma$ discrepant from the Planck value\cite{Cepheid}.
Since $N_{eff}$ and $H_0$ are {\em positively} correlated, a larger value for $H_0$ implies an increase in $N_{eff}$. In fact, the Planck collaboration has obtained $N_{eff} = 3.62^{+0.50}_{-0.48}$ when the larger value for $H_0$ is incorporated in the Planck data. Furthermore, when Baryon Acoustic Oscillation (BAO) data is taken into account, a similar value  of $N_{eff} = 3.54^{+0.48}_{-0.45}$ has been found. 

It is important to notice that, at present, measurements of $N_{eff}$ are in agreement with each other at the $1\sigma$ level.  The tension between direct $H_0$ measurements and the CMB and BAO data based on the standard $\Lambda$CDM paradigm can be relieved at the cost of additional neutrino-like species, as explicitly pointed out by the Planck Collaboration \cite{Planck}. Additionally, a recent analysis has been performed pointing to an evidence for dark radiation in the Planck data at $95\%$~C.L, if one takes into account at the same time observations of the CMB large angular scale polarization  from WMAP9 \cite{Planckanalyses}. Besides analyses including Planck data, recent studies involving the South Pole Telescope and ATACAMA telescope find an evidence for $N_{eff} > 3.04$ when data from different searches are taken into account \cite{Neffevidence}. 

In the present study, we seek to account for the tentative dark radiation component via partial non-thermal production of dark matter, which has been extensively investigated in the literature \cite{susynonthermal}, but just recently has arisen as an interesting scenario to reproduce the measured number of effective neutrinos \cite{Neffpaper0,Neffpaper}. In particular, Ref.~\cite{Neffpaper} has shown with a model independent approach that when a heavy particle decays into a WIMP-photon pair (where WIMP indicates a generic Weakly Interacting Massive Particle), the relativistic state for the non-thermally produced WIMPs could mimic the effect of one neutrino species at matter-radiation equality while evading Big Bang Nucleosynthesis (BBN) and structure formation bounds. 

Recent examples of explicit realizations of the non-thermal, relativistic WIMP scenario for the ``dark radiation'' include effective theories as well as a supersymmetric construction and many other models \cite{Neffpaper2,Neffpaper3}. Here, we show that such a scenario may also arise  with a weak-scale WIMP in the context of a non-supersymmetric 3-3-1 model, an electroweak extension of the Standard Model (SM) featuring a gauge group $ SU(3)_c\otimes SU(3)_L\otimes U(1)_N$. This model is a compelling alternative to the SM with a smoking gun signature given by the presence of charged gauge bosons and scalars, as well as a spectrum of particles whose phenomenological aspects have been investigated extensively \cite{331higgs,others}.  This model is also consistent with all electroweak bounds, while offering plausible explanations to many open problems in particle physics, such as dark matter \cite{331DM_1,331DM_2,331DM_3} and the number of particle generations \cite{families}.

In the 3-3-1 model we consider here, the dark matter particle is dominantly a thermally produced WIMP, which arises in the early universe via the standard thermal freeze-out picture, or ``WIMP miracle''. However, some fraction of the abundance of the dark matter particle has a non-thermal origin due to the decay of a right-handed singlet neutrino,  $N_R$,  which decays into WIMP-neutrino pairs, similar to the gravitino-sneutrino setup of certain supersymmetric models \cite{susysneutrino}. We also comment here on possible constraints on the injection of high energy neutrinos at early stages of the universe \cite{hadroniclimits}, and show that in our framework this is not a concern. Lastly, we show that the non-thermal production process  we invoke within our model is able to simultaneously reproduce the number of effective neutrinos measured by Planck and evade bounds from BBN, CMB and structure formation.

This study is organized as follows: in the next section we review the notion that WIMPs produced in a relativistic state can act effectively as ``dark radiation''; in the following section III we outline the particular particle physics setup we will hone in for the present analysis: the 3-3-1LHN model; section IV describes in general how dark radiation is realized in the context of 3-3-1LHN models, while section V examines in detail the relevant parameter space. Section VI, finally, we summarize and conclude.

\section{DARK MATTER PARTICLES AS DARK RADIATION}

In the standard $\Lambda$CDM picture, dark matter particles are non-relativistic at the time of structure formation. Nevertheless, if a fraction of the dark matter particles were produced with large enough kinetic energies, they would effectively behave as radiation, with their energy density being redshifted away until matter-radiation equality, i.e., quite similar to SM neutrinos. In order to determine the fraction and  energy density of the non-thermally produced dark matter particles allowed by BBN and structure formation bounds, we remind the Reader that at matter-radiation equality the energy density of one neutrino species is equal to $16\%$ of the total dark matter density. 
Hence, if  all the dark matter particles had an increase of $16\%$ in their boost factor, at matter-radiation equality, this would produce the same effect as one additional neutrino species. Of course this scenario where $100\%$ of the dark matter particles are produced relativistically is completely ruled out by structure formation. In other words, structure formation limits the fraction of dark matter particles produced with a large kinetic energy.

Concrete examples of this mechanism were studied in Ref.~\cite{Neffpaper2} for the case of a heavy particle decaying into a WIMP-photon pair. It was shown there that for suitable choices of the lifetime and daughter-to-mother mass ratio, such a mechanism would provide an interesting alternative to explain the currently mild evidence for $N_{eff}>3$ discussed above. More interestingly, a setup where some non-thermal production occurs after BBN is also a plausible explanation to why BBN and CMB probes indicate different values for the number of relativistic species ($N^{BBN}_{eff} \neq N^{CMB}_{eff}$). If the decay of the mother particle happens at a lifetime much greater than 100 seconds (BBN epoch) then no extra radiation would have been present during BBN. However, at the decoupling of the CMB, which happens at $\sim 10^{11}$~sec, some dark radiation could be detected due to the relativistic nature of some fraction of the dark matter particles. This effect is clearly pointed out in Figs. 1-2 of Ref.~\cite{Neffpaper}.

``WIMPy dark radiation'' is thus a potentially successful explanation to the tentative evidence for additional relativistic degrees of freedom in the early universe. A non-thermal production setup would be devastating for damping the evolution of structures at small scales, having an impact similar to hot dark matter if a significant fraction of the dark matter particles were indeed produced with a large kinetic energy. 
Quantitatively, it has been  shown that in order to be consistent with structure formation bounds, at most roughly $1\%$ of  all dark matter particles might have had a non-negligible kinetic energy at matter-radiation equality \cite{Neffpaper}. Moreover, BBN bounds are quite stringent as well, imposing limits on the energy released and on the lifetime of the mother particle. When the particle produced along with the WIMP interacts mostly electromagnetically (such as a photon or electron), it has been concluded that for lifetimes shorter than $10^4$~seconds, BBN bounds are evaded and structure formation limits are circumvented as long as at most $1\%$ of the dark matter particles are produced with large kinetic energies (see Fig.~1 of Ref.\cite{Neffpaper2}).

Turning again our attention to the relation between the non-thermal production of dark matter particles and the number of effective neutrinos. In this work we will derive this relation following closely the procedure in Ref.~\cite{Neffpaper2}. There, in the scenario where some fraction ($f$) of the dark matter particles are produced along with neutrinos via general decay, $X^{\prime} \rightarrow WIMP + \nu$, the dark radiation mimicked by those dark matter particles at the matter radiation equality reads,
\begin{eqnarray}
\Delta N_{eff} & \simeq & 4.87 \times 10^{-3}\left( \frac{\tau}{10^6\ s} \right)^{1/2}\nonumber\\ 
                     &   &
  \times \left[\left( \frac{ M_{X^{\prime}} }{2M_{\rm wimp}} + \frac{M_{\rm wimp}}{2M_{X^{\prime}}} -1 \right) \right]\times f.
\label{deltaNeff2}
\end{eqnarray}
However, this equation is valid only in the ultra-relativistic limit and $10\%$ error is generated compared to the fully relativistic equation. This error is due to an approximation used in the boost factor. In general, the boost factor of the dark matter particles at given time is given by,
\begin{equation}
\gamma^2_{DM} = \left(\frac{a_{\tau}}{a}\right)^{2} \left( (\gamma^{\tau}_{DM})^2 -1 \right) + 1 .
\label{proof11}
\end{equation}where,
\begin{equation}
\gamma_{DM}^{\tau} = \left( \frac{M_{X^{\prime}} }{ 2M_{DM}} + \frac{M_{DM}}{ 2M_{X^{\prime}}} \right).
\label{proof1}
\end{equation}which is the boost factor at the decay, and $a_{\tau}/a_{eq} = 7.8 \times 10^{-4} (\tau/10^6\ s)^{1/2}$. 
Therefore using the fact that $\Delta N_{eff} = f\left( \gamma_{DM} -1 \right)/0.16$, we get,
\begin{eqnarray}
\Delta N_{eff} & = & \frac{f}{0.16}\left[ \left( \sqrt{7.8^2\cdot 10^{-8}\left(\frac{\tau}{10^6\ s}\right) \left( (\gamma^{\tau}_{DM})^2\right) + 1 }\right) - 1 \right].\nonumber\\ 
\label{proof14}\nonumber\\
\end{eqnarray}in the limit that $\gamma^{\tau}_{DM} \gg 1$. 
In summary, Eq.(\ref{proof14}) determines the number of effective neutrino mimicked by the non-thermal production of dark matter particles and we will be using this Eq.(\ref{proof14}) throughout the paper.

\begin{figure}[]
\centering
\includegraphics[width=1\columnwidth]{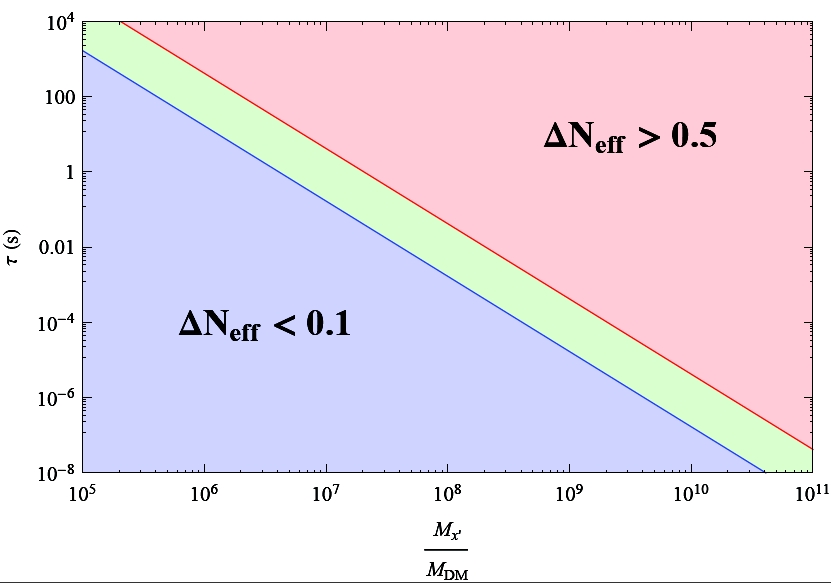}
\caption{Region of the parameter space {\it lifetime ($\tau$)} $\times$ {\it mass ratio ($ M_{X^{\prime}}/M_{\rm wimp}$) } which reproduces $\Delta N_{eff} \leq 0.1\ (\mbox{blue})$, $0.1 \leq \Delta N_{eff} \leq 0.5\ (\mbox{green})$, $\Delta N_{eff} \geq 0.5\ (\mbox{pink})$.  }
\label{figure1}
\end{figure}
In Fig.~\ref{figure1} we show contours for $\Delta N_{eff}$ in the {\it $\tau$} $\times$ {\it $M_{X^{\prime}}/M_{\rm wimp}$} parameter space where $\Delta N_{eff} \leq 0.1\ (\mbox{blue})$, $0.1 \leq \Delta N_{eff} \leq 0.5\ (\mbox{green})$, $\Delta N_{eff} \geq 0.5\ (\mbox{pink})$. It is important to emphasize that Eq. (\ref{deltaNeff2}) does not depend on the nature of the particles involved and therefore it is valid for any decay mode, as long as the mass of the mother particle is significantly heavier than the mass of the decay products, and of the stable WIMP in particular. Therefore, in principle, any particle physics model that contains a long lived particle decaying into dark matter particles after WIMP freeze-out might induce $\Delta N_{eff}\neq 0$. Furthermore, Eq.~(\ref{deltaNeff2}) shows that if $1\%$ of the whole dark matter of the universe was produced by the decay of a heavy particle with a lifetime equal or shorter than $\tau=10^4$~sec one needs to have a mass, for the heavy mother particle such that
\begin{equation}
\label{eq:ratio}
\frac{M_{X^\prime}}{M_{\rm wimp}}\gtrsim4\times 10^5\ \Delta N_{eff},
\end{equation}in agreement with \cite{Neffpaper,Neffpaper2,Neffpaper3}.
This implies, in particular, that in order to explain the Planck result of $\Delta N_{eff}\simeq 0.62$ when including the direct measurements of $H_0$ from HST, we need  $M_{X^\prime}/M_{wimp} > 2.5 \times 10^5$. In other words, the mother particle must be {\em significantly heavier} than its decay products. It is worth reiterating that, in this framework, the majority of the dark matter particles would still have to be produced as cold, i.e. non-relativistic particles, presumably with a thermal cross section at the electroweak scale providing the right thermal relic abundance. Our goal here is to investigate if this scenario is feasible in a well-motivated electroweak gauge group extension of the SM  named 3-3-1LHN which we  briefly introduce below.

\section{The 3-3-1LHN Model}
\label{sec1}
3-3-1 models  refer to  electroweak extensions of the SM gauge group based on the enlarged gauge group $ SU(3)_c\otimes SU(3)_L\otimes U(1)_N$ that lies at $\sim$ TeV scale. 3-3-1 models potentially address important theoretical and phenomenological questions which remain unexplained within the SM, such as the number of particle families \cite{families}, certain dark matter signals \cite{331DM_1,331DM_2}, the possible Higgs to diphoton excess \cite{331higgs}, electric charge quantization\cite{ecq}, etc. In addition, 3-3-1 models present a rich phenomenology which includes new scalars and gauge bosons, as extensively explored in the literature \cite{others}. For these and many other reasons, 3-3-1 models stand as compelling alternatives to the SM.  It is worth also to remark that interesting proposals have been put forth recently concerning dark matter in 3-3-1 gauge symmetries, see e.g. Refs.~\cite{Dong}. Here, however, we focus on a version of this class of models we indicate as 3-3-1LHN \cite{model331}, which has two noticeable distinct features compared to previous versions \cite{families,minimalversion}: 

(1) the presence of neutral fermions, $N_L$ and $N_R$, and 

(2) a scalar field as a dark matter candidate. 

\noindent We briefly introduce the 3-3-1LHN model in the following sections.

\subsection{Particle content}

In the 3-3-1LHN model, which has a scale of symmetry breaking at $\sim 1$~TeV, the left-handed standard leptons and the neutral fermion $N_L$ compose a triplet of $SU(3)_L$, $L_a=(\nu_{aL},l_{aL},N_{aL})^T$,  while right-handed leptons come in singlets, $e_{a R}\,,\,N_{aR}$, where the subscript {\it a} runs over the three generations. A key distinction between the model developed here and the one in Ref. \cite{model331} is that the extra neutral fermions, $N_{L,R}$ do not carry lepton number, eliminating the need of any bilepton in the model, as was the case for the extra quarks, some scalars
and gauge bosons in the previous 331LHN. This is required by the discrete symmetry that  guarantees the stability of our dark matter candidate.  As for the hadronic sector, the first two families of left handed fields are arranged in anti-triplet representations, $Q_{i L} = (d_{i L}, -u_{i L}, q^{\prime}_{i L} )^T$ with $i=1,2$, and the third in a triplet representation, with $Q_{3 L} = (u_{3 L}, d_{3 L}, q^{\prime}_{3 L} )^T$.  Concerning right-handed quarks, they are all singlets, with hyper-charges exactly equal to their electric charges. Notice that the three quarks ($q^{\prime}$) shown above are new quarks added to the Standard Model.  Three triplets of scalars,  $\chi  =  (\chi^0, \chi^-, \chi^{\prime 0})^T, \rho =  (\rho^+, \rho^0, \rho^{\prime +})^T, \eta  =  (\eta^0, \eta^-, \eta^{\prime 0})^T$, are necessary  to induce the proper pattern of symmetry breaking after they develop vacuum expectation values (VEVs) different from zero, 
\begin{eqnarray}
 \eta^0 , \rho^0 , \chi^{\prime 0} \rightarrow  \frac{1}{\sqrt{2}} (v_{\eta ,\rho ,\chi^{\prime}} 
+R_{ \eta ,\rho ,\chi^{\prime}} +iI_{\eta ,\rho ,\chi^{\prime}})\,,
\label{vacua} 
\end{eqnarray}
and then generate, at tree level,  masses for all massive particles in the model. Besides the standard gauge bosons, $W^{\pm}\,\, Z^0$ and the photon, the model contains five new gauge bosons indicated as  $V^{\pm}\,,\,U^{0}$, $U^{0 \dagger}$ and a $Z^{\prime}$ from the enlarged gauge group.

\subsection{Scalar spectrum and Mass Eigenstates}

We have invoked a  R-parity discrete  symmetry quite similar to the one in the minimal supersymmetric standard model case, which we indicate with
$P=(-1)^{3(B-L)+2s}$, where $B$ is the baryon number, $L$ is the lepton number and $s$ is spin of the field. Thus, we have  the following assignments of {\it P}  carried by the particle content:
\begin{eqnarray}
(N_L\,,\,N_R\,,\,d^{\prime}_i\,,\,u^{\prime}_3\,,\,\rho^{\prime +}\,,\,\eta^{\prime 0}\,,\,\chi^{0}\,,\,\chi^-\,,\, V\,,\,U) \rightarrow -1.
\label{R-P}
\end{eqnarray}where $d^{\prime}_i$ and $u^{\prime}_3$ are new heavy quarks predicted in the model due to the enlarged gauge group. The remaining fields all transforming trivially under this symmetry. The lightest neutral particle odd by R-parity  symmetry is, in principle, a viable dark matter candidate. We will see that it will be a linear combination of the neutral scalars $\chi^{0}$ and $\eta^{\prime 0 *}$ .

We will ignore the charged scalars, gauge bosons, as well as the heavy quarks in the model, since they do not play any role throughout this work. They are assumed to be heavy with their masses proportional to the scale of symmetry breaking of the model.

The R-parity symmetry  allows us to write the most general scalar potential and Yukawa Lagrangian, respectively, as:

\begin{eqnarray} & V(\eta,\rho,\chi) & = \mu_\chi^2 \chi^2 +\mu_\eta^2\eta^2
+\mu_\rho^2\rho^2+\lambda_1\chi^4 +\lambda_2\eta^4
+\lambda_3\rho^4 \nonumber \\
                                   &   &+\lambda_4(\chi^{\dagger}\chi)(\eta^{\dagger}\eta)
+\lambda_5(\chi^{\dagger}\chi)(\rho^{\dagger}\rho)+\lambda_6
(\eta^{\dagger}\eta)(\rho^{\dagger}\rho) \nonumber \\
                                   &   &+\lambda_7(\chi^{\dagger}\eta)(\eta^{\dagger}\chi)
+\lambda_8(\chi^{\dagger}\rho)(\rho^{\dagger}\chi)+\lambda_9
(\eta^{\dagger}\rho)(\rho^{\dagger}\eta) \nonumber \\
                                   &   &-\frac{f}{\sqrt{2}}\epsilon^{ijk}\eta_i \rho_j \chi_k +\mbox{h.c.},
\label{potential}
\end{eqnarray}

\begin{eqnarray}
&-&{\cal L}^Y =f_{ij} \bar Q_{iL}\chi^* d^{\prime}_{jR} +f_{33} \bar Q_{3L}\chi u^{\prime}_{3R} + g_{ia}\bar Q_{iL}\eta^* d_{aR} \nonumber \\
&&+h_{3a} \bar Q_{3L}\eta u_{aR} +g_{3a}\bar Q_{3L}\rho d_{aR}+h_{ia}\bar Q_{iL}\rho^* u_{aR} \nonumber \\
&&+ G_{ab}\bar f_{aL} \rho e_{bR}+g^{\prime}_{ab}\bar{f}_{aL}\chi N_{bR}+ \frac{M}{2} \bar{N_{bR}^c} N_{bR} + \mbox{h.c.}.\nonumber \\
\label{yukawa}
\end{eqnarray}
The  masses of the new neutral fermions are given by the last two terms of Eq.~(\ref{yukawa}). Without the last term those particles would have Dirac masses at the TeV scale. However, with the inclusion of the last Majorana mass term, one obtains masses set by a see-saw type I mechanism which can be much larger than the TeV scale, as needed to obtain a large daughter-to-mother mass ratio. We stress here that the role played by these new neutral fermions are twofold: to give rise to a seesaw mechanism and to generate the non-thermal production of dark matter. Hence $N_R$'s are heavy particles which decouple from the rest of the 3-3-1 particle spectrum.  In this case, this mass term does not affect the stability of our WIMP. We call attention to the fact that another difference among this model and the one in Refs. \cite{331DM_2} is  the bare mass terms for the  $N_R$'s.

The Yukawa interactions of Eq.~(\ref{yukawa}) above provide Dirac mass terms for all charged fermions in the 3-3-1LHN. The standard neutrinos, $\nu_L$'s, gain Majorana mass terms through effective operators, as described in Ref.\cite{331nu}. On the other hand, the heavy neutral fermions, $N_{L,R}$'s, acquire Majorana mass terms through a kind of type I see-saw  mechanism engendered by the two last terms in the Yukawa interactions as described below.

To understand the see-saw mechanism, notice that the last two terms in Eq.(\ref{yukawa}) give rise to the following mass matrix in the basis ($N_L\,,\,N_{R}^C$):
\begin{equation}
\left(\begin{array}{cc}
0 & m_D\\
m_D & M\\
\end{array}\right),
\end{equation}
where $m_D= g^{\prime}_{ab}v_{\chi^{\prime}}$, $N_L=(N_{1L}\,,\,N_{2L}\,,\,N_{3L})$ and $N^C_{R}= (N_{1R}^C\,,\,N_{2R}^C\,,\,N_{3R}^C)$. Diagonalizing the matrix above gives
\begin{equation}
N^{\prime}_L = N_L + \frac{m_D}{M} N_R^c \ \,\,\,\mbox{and}\ N^{\prime}_R = N_R + \frac{m_D}{M} N_L^c,
\end{equation}with
\begin{equation}
M_{N^{\prime}_L} = \frac{m_D^2}{M}\ \mbox{and}\ M_{N^{\prime}_R} = M.
\label{massesSEESAW}
\end{equation}
Therefore in the limit $M \gg m_D$, which is required in our setup, we find $N^{\prime}_L  \simeq N_L$ and $N^{\prime}_R \simeq N_R$ with mass $\frac{m_D^2}{M}$ and $M$, respectively. For the sake of simplicity, we assume throughout this work that $M_D$ and $M$ are diagonal. Such mixing among the heavy fermions $N_L$ and $N_R$ gives rise to an interaction $g^{\prime}\frac{m_D}{M}\bar \nu_L \phi N^{\prime C}_L $ which will not affect the WIMP stability as long as $\phi$ is assumed to be the lightest particle in the spectrum, as we enforce here to be the case. This condition will turn out to be rather restrictive to our model as we will see in Figs.\ref{g1_large1}-\ref{g1_small}.  

Concerning the scalar  mass spectrum of the 3-3-1LHN, our model supplements the SM by adding two CP-even scalars, $S_1$  and $S_2$,  with masses given by,
\begin{eqnarray}
M_{S_{1}} & = & \sqrt{ \frac{v^{2}}{4}+2\lambda_{1}v_{\chi^\prime}^{2}}\,, \nonumber \\
M_{S_{2}} & = &\sqrt{ \frac{1}{2}\left( v_{\chi^\prime}^{2}+2v^{2}(2\lambda_{2}-\lambda_{6})\right)}\,, 
\label{massashiggs}
\end{eqnarray}
while the standard Higgs, $H$,  has mass given by $M_{H}  =  \sqrt{ 3\lambda_{2}}v $, where $v^2_\eta + v^2_\rho = v^2=(246\mbox{GeV})^2$ (in this work we assume $v_\eta=v_\rho$). The corresponding eigenstates are given by
\begin{eqnarray}
S_1 =R_{\chi^{\prime}} \,, S_2 = \frac{(R_\eta - R_\rho)}{\sqrt{2}}\,,H = \frac{(R_\eta +R_\rho)}{\sqrt{2}}.
\end{eqnarray}
The model also features  a CP-odd scalar with mass given by $
M_{P_{1}} =\sqrt{ \frac{1}{2}(v_{\chi^\prime}^{2}+\frac{v^{2}}{2}) }$, and a complex neutral scalar, which is the WIMP candidate we consider here and which we indicate with the symbol $\phi$, with $\phi \approx  v/v_{\chi^\prime}\ \chi^{0\star} + \eta^{\prime 0}$, featuring a mass
\begin{eqnarray}
M_{\rm wimp} & = &\sqrt{ \frac{(\lambda_{7} + \frac{1}{2} )}{2}[v^{2}+v_{\chi^\prime}^{2}] }.
\label{massfi}
\end{eqnarray}

The scalar $\phi$  will be chosen as the lightest odd R-parity particle. Thus it is the cold dark matter particle in the present setup. As shown in Ref.~\cite{GC331} (see in particular Fig.~3), such particle can naturally provide the correct relic abundance and be consistent with current direct detection bounds \cite{GC331}.  Additionally, the $\phi$ particle can also in principle explain the gamma-ray excess in the Galactic Center  observed in the Fermi-LAT satellite data \cite{GCpaper}. This scalar is a WIMP, whose relic abundance is mainly thermally produced in the early universe via interactions with SM particles. It is important to stress that $\phi$ has also a non-thermal component, however, because the heavy fermion $N_R$ may decay  into  $\phi\ \nu$ pairs, as allowed by the last term in the Yukawa interaction Lagrangian of Eq.~(\ref{yukawa}). 

There are two charged scalars in the spectrum ($h_1$ and $h_2$) with masses linearly proportional to the scale of symmetry breaking of the model ($v_\chi^{\prime}$). These scalars are not relevant in this work and will be ignored. With respect to the gauge bosons,  the masses of the five extra gauge bosons are  given by, 
\begin{eqnarray}
m^2_{V}    &=& m^2_{U^0} = \frac{1}{4}g^2(v_{\chi^\prime}^2+v^2)\,,
\nonumber \\
m^2_{Z^\prime} &=& \frac{g^{2}}{4(3-4s_W^2)}[4c^{2}_{W}v_{\chi^\prime}^2 +\frac{v^{2}}{c^{2}_{W}}+\frac{v^{2}(1-2s^{2}_{W})^2}{c^{2}_{W}}]\,,\nonumber\\
\label{massvec}
\end{eqnarray}where $V^{\pm}$ are charged gauge bosons which mimic the couplings of SM gauge boson W, and $U^0$ is a  complex neutral gauge boson. We note that these gauge bosons provide a smoking gun signature for 3-3-1 models. However, similar to the aforementioned charged scalars ($h_1$ and $h_2$), these bosons will not be important in the reasoning developed here and will thus be ignored hereafter.

To summarize, the 3-3-1LHN model has in its spectrum a scalar WIMP dark matter candidate, $\phi$,  which  provides most of the observed cold dark matter through standard freeze-out (see Fig.~4 of Ref.~\cite{331DM_2} and Fig.~1 of Ref.~\cite{GC331}) and a spin-independent scattering cross section off nuclei consistent with current limits, as well as heavy fermions ($N_{R}$). The lightest of these fermions ($N_{1R}$) will play a major role in our results as we shall see further, acting as the ``mother particle'' for the small relativistic population of $\phi$'s responsible for the dark radiation component. 

\section{Dark Radiation in the 3-3-1LHN Model: General Considerations}
\label{darkradiation}
There are few requirements for the ``WIMPy'' dark radiation scenario to be realized in a given particle physics model, namely:

I. The mass of the mother particle ($N_{R}$) must be much greater than the mass of the WIMP ($\phi$), according to Eq.(\ref{eq:ratio});

II. The lifetime of the mother particle should be shorter than $10^4$ sec to circumvent BBN bounds;

III. Just a small fraction ($\sim 1\%$ or smaller) of the dark matter particles ($\phi$) should be produced via this non-thermal mechanism, in order not to spoil structure formation.

%
%
%
We stated in Eq.(\ref{eq:ratio}) $M_{X^\prime} \gtrsim4\times 10^5\ \Delta N_{eff} M_{\rm wimp}$. Therefore, this non-thermal production mechanism is only able to mimic the effect of one neutrino species when the mother particle is much heavier than the WIMP. For instance, for a $100$~GeV WIMP, $M_{X^\prime} \geq 4 \times 10^6$ for $\Delta N_{eff}=0.1$. This large mass ratio leads to a crucial fact that should be highlighted. Since the WIMP inherits the abundance of the mother particle, i.e. $\Omega_{N_{1R}}=M_{N_{1R}}/M_{\rm wimp} \Omega_{\phi}$, where $\Omega_{\phi}$ is the relative abundance of $\phi$ coming from the decay of $N_{1R}$, we find that \footnote{Eq.(\ref{OmegaN1R_1}) is valid because we are matching the abundances at the matter-radiation equality, when the dark matter particles produced relativistically have become essentially non-relativistic due to the expansion of the Universe.}, 

\begin{equation}
\Omega_{N_{1R}}=M_{N_{1R}}/M_{\rm wimp}\cdot f\cdot \Omega_{DM}.
\label{OmegaN1R_1}
\end{equation}
We can use Eq.(\ref{eq:ratio}) to find,

\begin{equation}
\Omega_{N_{1R}} \gtrsim 4\cdot 10^5\ \Delta N_{eff}\cdot f\cdot \Omega_{DM},
\label{OmegaN1R_2}
\end{equation}

\begin{figure}[]
\centering
\includegraphics[scale=0.5]{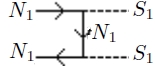}
\includegraphics[scale=0.7]{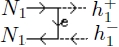}
\includegraphics[scale=0.7]{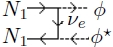}
\includegraphics[scale=0.7]{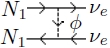}
\includegraphics[scale=0.7]{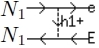}
\caption{Annihilation channels that contribute to the abundance of $N_{1R}$. $\phi$ is the WIMP of our model, $h_1^+$ is a singly charged scalar, and $S_1$ a CP even scalar.}
\label{fig2new}
\end{figure} 

The $N_{1R}$ abundance froze-out much earlier than the decay. So $\Omega_{N_{1R}}$ is the abundance of $N_{1R}$ as if it had not decayed. In Fig\ref{fig2new} we show the Feynman diagrams that contribute to the abundance of $N_{1R}$, which was computed using Micromegas \cite{micromegas}. Its abundance scales with $g_{11}^{\prime -4}$. Therefore as we decrease the $g_{11}^{\prime}$ coupling the abundance goes up quickly and an entropy dilution mechanism that we will discuss further will be needed, in order not to overproduce non-thermal WIMPs. Taking $\Delta N_{eff}=0.1,f=0.01$ and $\Omega_{DM}\sim 0.23$ we find that $\Omega_{N_{1R}} \gtrsim  10^3$.  In other words the abundance has to be much greater than one at decay. This is an important point because when we later compute the abundance of the mother particle as a function of its mass and of the coupling $g^{\prime}_{11}$ using the Micromegas package \cite{micromegas}, we will be able to directly  reconstruct the mass of the WIMP using Eq.(\ref{OmegaN1R_1}) once we fix $f=0.01$ and $\Omega_{DM}\sim 0.23$, and then check what is the associated number of effective neutrinos induced by the chosen setup, as will be shown in Figs.\ref{g1_large1}-\ref{g1_small}.

The dark radiation model studied here has no effect on the abundance and spin independent cross section of the WIMP, because the only parameter involved in both is the mass of the WIMP. Hence the dark radiation setup can be investigated in parallel with no prejudice concerning the WIMP miracle, which can be realized for a wide range of WIMP masses, as shown in \cite{331DM_1,331DM_2}. However, if there is a long-lived particle that decays after the WIMP freezes-out, which happens at temperatures of $M_{\rm wimp}/20$ ($10^{-8}$~s for a $100$~GeV WIMP), some small fraction of the dark matter particles will be non-thermally produced and might behave as ``dark radiation''. As long as this fraction is small, of order of $1\%$ or less, this non-thermal production mechanism is completely consistent with structure formation bounds, as shown in Refs.~\cite{Neffpaper,Neffpaper2}.  

We would like to point out that in the limit of large hierarchy, the Lorentz factor for the dark matter daughter particle is $\gamma\sim M_{X^\prime}/M_{\rm DM}$ and this Lorentz factor is suppressed by, approximately, $a_{eq}/a_\tau\sim10^4(\tau/10^4 s)$, thus it is not inconceivable that for the largest lifetimes and mass ratios the dark matter be relativistic at matter-radiation equality. However, unlike what originally put in the manuscript, we use Eq.(14) at late times for the purpose of matching the dark matter abundance observed today, and not at matter-radiation equality.

\begin{figure}[t]
\centering
\includegraphics[width=1\columnwidth]{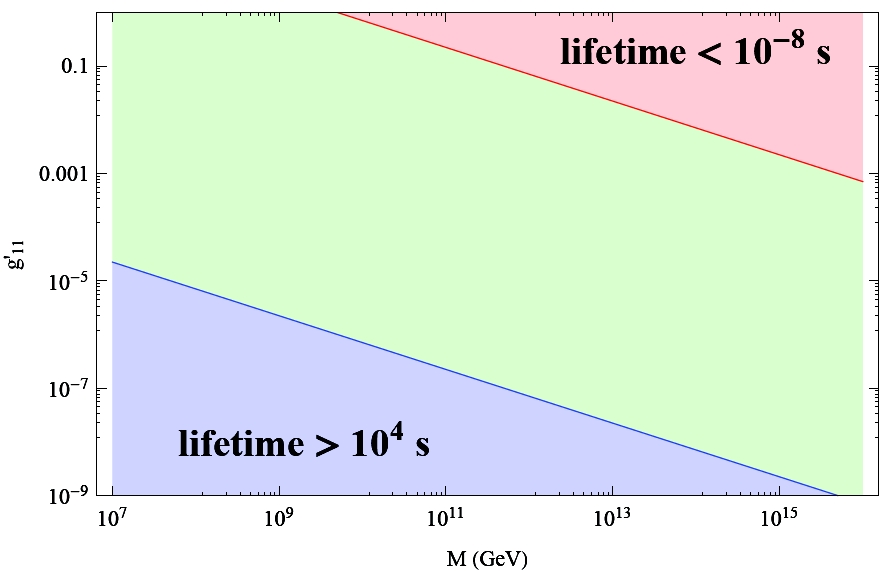}
\caption{Region of the parameter space $M$ $\times$ $g^{\prime}_{11}$ allowed by BBN ($\tau \leq 10^4$~s) and with decay that occurs after the WIMP freeze-out ($\tau \geq 10^{-8}$~s) for $v_{\chi}^{\prime}=1$~TeV is shown in green. }
\label{lifetime331}
\end{figure}

The BBN constraint on the lifetime of the mother particle in general depends on what is produced in the final state, the total energy injected and the branching ratio. Here we have neutrinos in the final states, therefore one might expect weaker constraints as oppose to the pure electromagnetic case which requires the lifetime to be shorter than $10^4$~s. Nevertheless, 3 and 4-body hadronic decays might be induced with smaller branching ratios as described in Ref.\cite{hadroniclimits}. In particular, the latter bounds depend on the injected energy and the branching ratio into hadronic states. In this work, we are being conservative and for this reason we assumed use the limit obtained when we have a photon in the final state.  In this work we were trying to investigate the validity of this dark radiation scenario in this model and we believe that the derivation of the BBN bounds for neutrinos in the final states is out of the scope of this work. Therefore in summary we will assume that the lifetime of the $N_{1R}$ has to be shorter than $10^4$~s due to BBN constraints. The most important parameters which control the lifetime of this heavy fermion are the scale of symmetry breaking of this model, its mass and the Yukawa coupling, $g^{\prime}_{11}$. Therefore, it is important to compute the partial width to the dominant decay mode $N_{1R}\rightarrow WIMP + \nu_e$, which reads
\begin{equation}
\Gamma(N_{1R}\rightarrow WIMP+ \nu_e) = \frac{\lambda^2}{64\pi} \left(\frac{M}{1\ \mbox{GeV}}\right) \left( 1- \frac{M_{\phi}^2}{M^2}\right)^2,
\end{equation}where $\lambda = g^{\prime}_{11}v/v_{\chi^{\prime}}$ and M is the see-saw scale. The lifetime in this case is given by
\begin{equation}
\tau \simeq  \left(5\cdot10^{-5}{\rm sec}\right)\left(\frac{10^{-3}}{g^{\prime}_{11}}\right)^2 \left(\frac{v_{\chi^{\prime}}}{10^3\ \mbox{GeV}}\right)^2\left( \frac{10^{12}\ \mbox{GeV}}{M}\right) .
\label{lifetime}
\end{equation}
From Eq.(\ref{lifetime}) we notice that there will be a very wide range of Yukawa couplings ($g^{\prime}_{11}$) that produce lifetimes in the range allowed by BBN ($\tau \leq 10^4$~s) and with decays that occur after the WIMP freezeout ($\tau \geq 10^{-8}$~s). In Fig.~\ref{lifetime331} we show in green the region of the parameter space $M$ $\times$ $g^{\prime}_{11}$ allowed, for $v_{\chi}^{\prime}=1$~TeV. 

%
%
%
\begin{figure}[t]
\centering
\includegraphics[width=1\columnwidth]{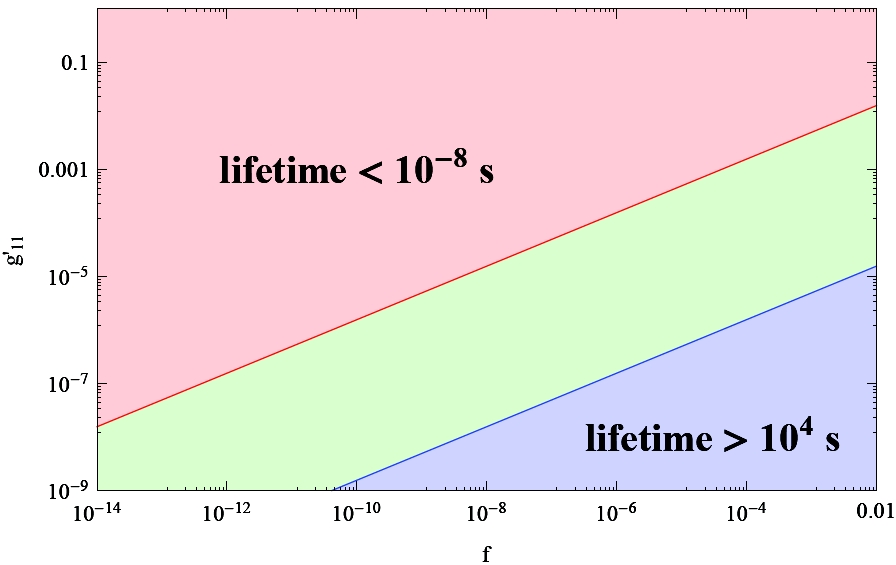}
\caption{The allowed ranges in the $g_{11}^{\prime},f$ plane with lifetimes in the correct ranges are again shown in green for $M_{WIMP}=100~$GeV, $v_{\chi}^{\prime}=1$~TeV and $\Delta N_{eff}=0.5$. }
\label{BBN}
\end{figure}
We can substitute the two expressions for the lifetime given in  Eq.~(\ref{lifetime}) into Eq.~(\ref{deltaNeff2}) to eliminate one of the  free parameters.  For example, if we eliminate the heavy fermion mass we find the following expression for the WIMP mass:
\begin{eqnarray}
\label{lowerbound}
M_{\rm wimp} & \leq & 100\ \mbox{GeV}\cdot \left( \frac{10^{-13}}{g_{11}^{\prime}}\right)\left(\frac{ v_{\chi}^{\prime}}{10^3\ \mbox{GeV}} \right)^{1/2}\times \nonumber\\
&  & \left(\frac{M}{2\cdot 10^{8}\ \mbox{GeV}}\right)^{1/2} \left( \frac{f }{0.01}\right)  \left( \frac{0.5}{\Delta N_{eff}}\right).
\end{eqnarray}
The required range for the lifetime produces the allowed region in the $g_{11}^{\prime},f$ plane we show in Fig.~\ref{BBN}.  In this case, we have chosen $M_{\rm wimp}=100~$GeV, $v_{\chi}^{\prime}=1$~TeV and $\Delta N_{eff}=0.5$.  We thus find a very wide range of parameters which can in principle reproduce $\Delta N_{eff}=0.5$ with a 100~GeV WIMP. In other words, a $100$~GeV WIMP is perfectly capable of mimicking the additional effective half neutrino species in the early Universe.  As a side comment, we also notice that as the fraction of dark matter particles that are produced non-thermally is decreased, the amount of fine-tuning required in the Yukawa coupling rapidly increases.

\section{A WIMPy Dark Radiation 3-3-1 Model}

\begin{figure}[!t]
\centering
\includegraphics[width=1\columnwidth]{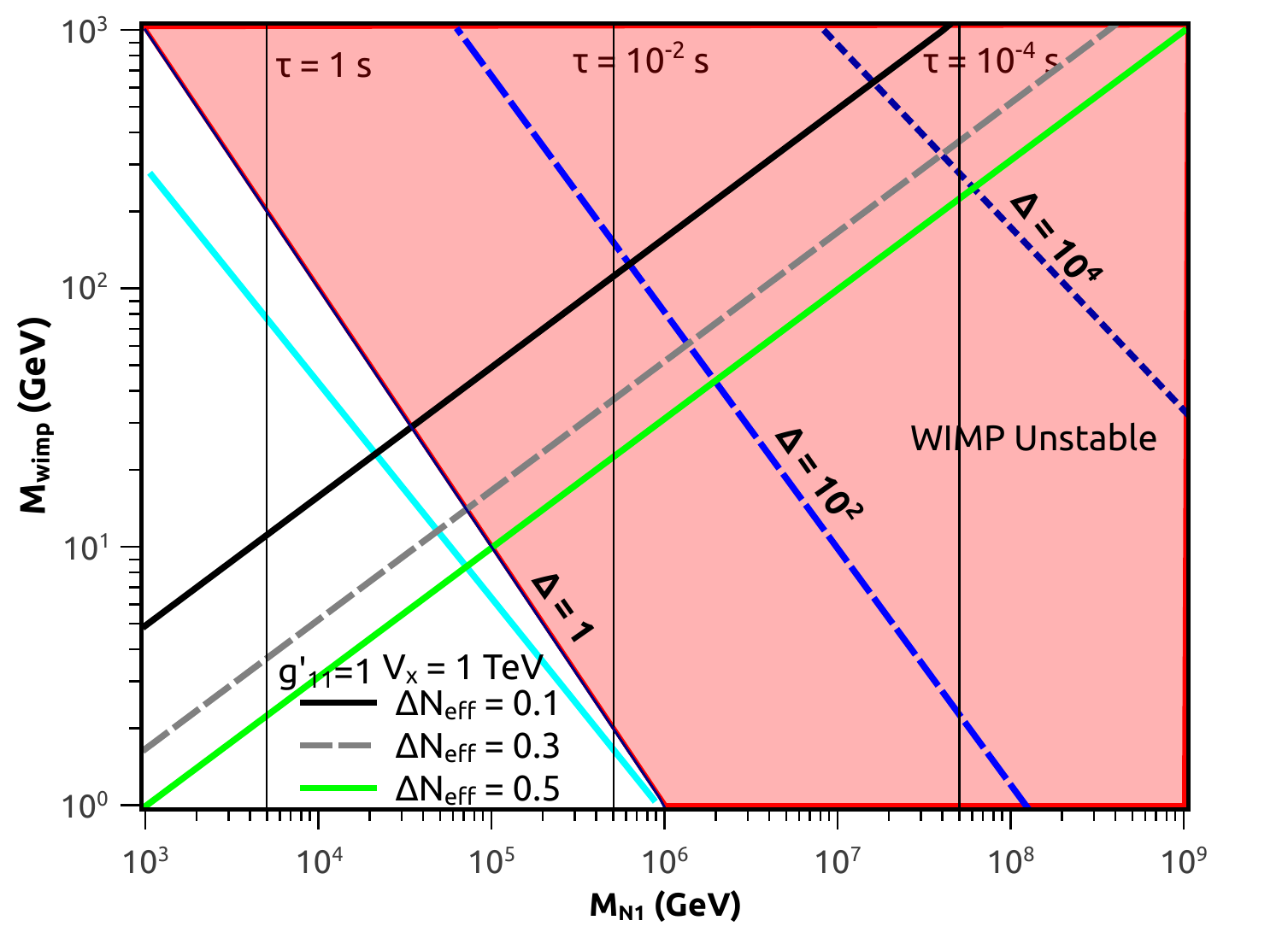}
\caption{The ``mother-daughter'' particle mass parameter space, for $g_{11}^\prime=1$. The red shaded region induces the WIMP decay. The vertical lines indicate constant values of the mother particle lifetime. The diagonal lines indicate the induced variation in the number of effective relativistic degrees of freedom $\Delta N_{\rm eff}$ and the entropy dilution factor $\Delta$ needed to suppress the mother particle relic density. The cyan $\Delta=1$ line corresponds to standard cosmology without any entropy dilution needed.}
\label{g1_large1}
\end{figure}

\begin{figure}[!t]
\centering
\includegraphics[width=1\columnwidth]{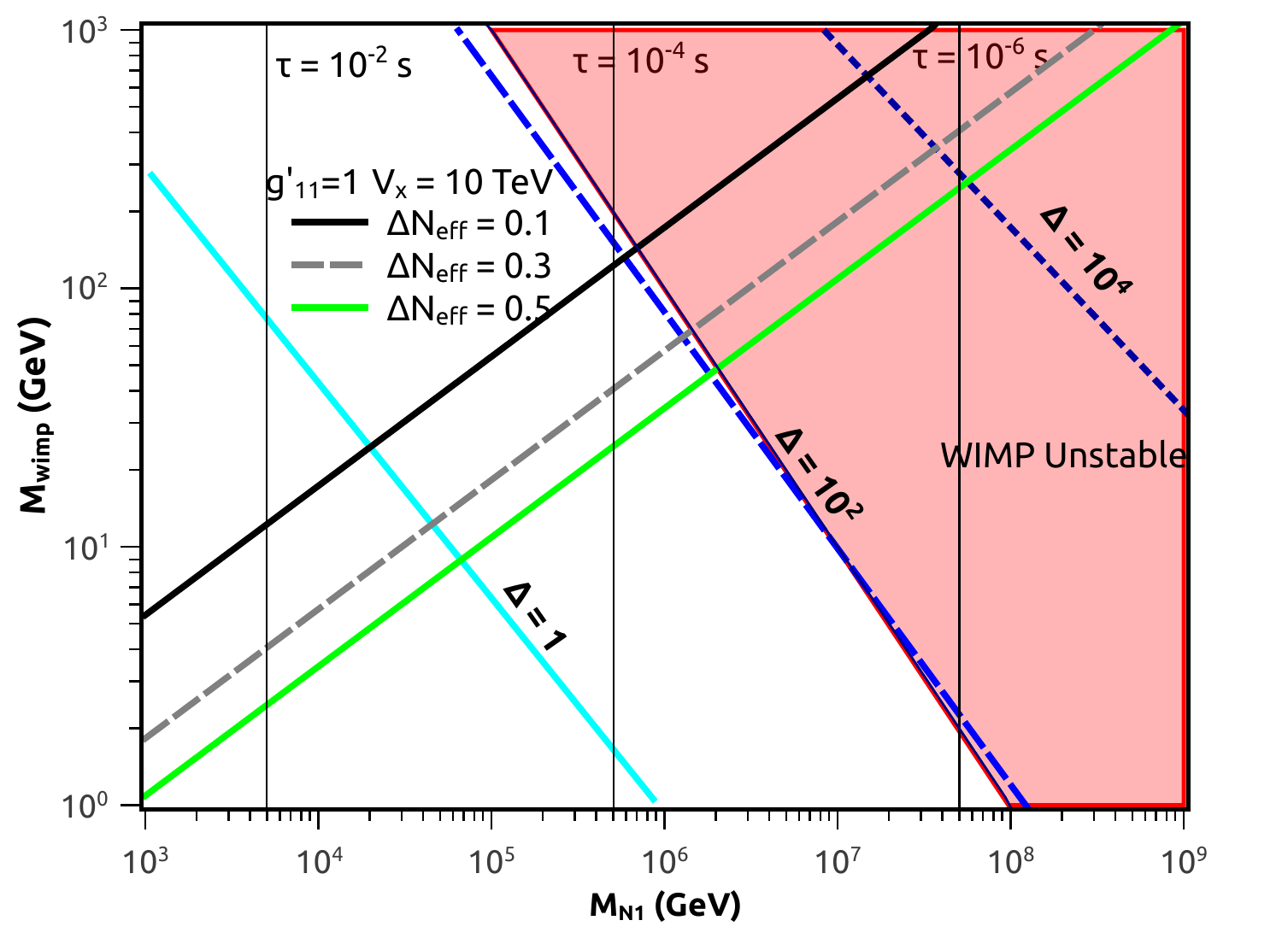}
\caption{The ``mother-daughter'' particle mass parameter space, for $g_{11}^\prime=1$. The red shaded region induces the WIMP decay. The vertical lines indicate constant values of the mother particle lifetime. The diagonal lines indicate the induced variation in the number of effective relativistic degrees of freedom $\Delta N_{\rm eff}$ and the entropy dilution factor $\Delta$ needed to suppress the mother particle relic density. The cyan $\Delta=1$ line corresponds to standard cosmology without any entropy dilution needed.}
\label{g1_large2}
\end{figure}

\begin{figure*}[!t]
\centering
\mbox{\includegraphics[width=0.95\columnwidth]{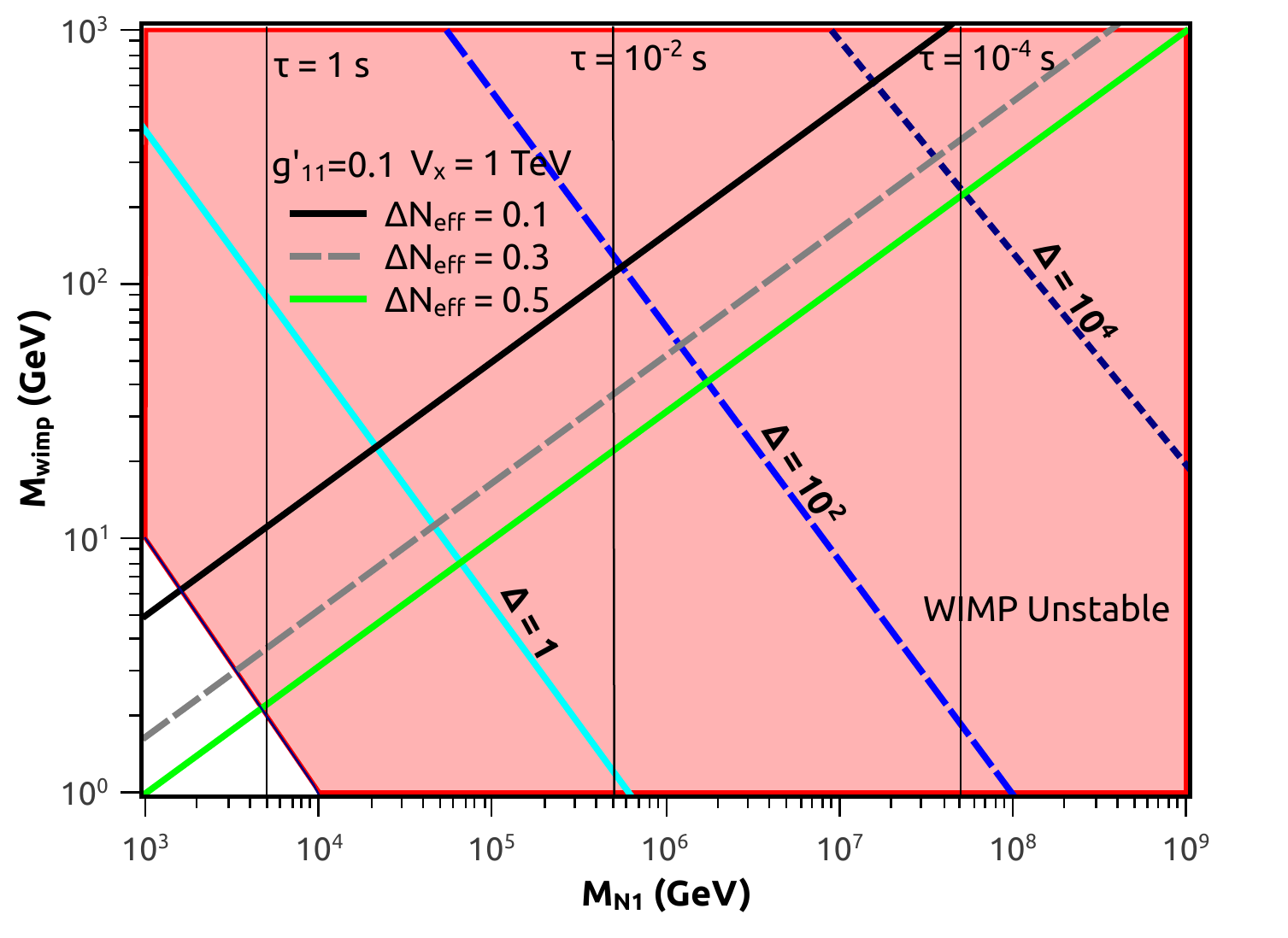}\quad\includegraphics[width=\columnwidth]{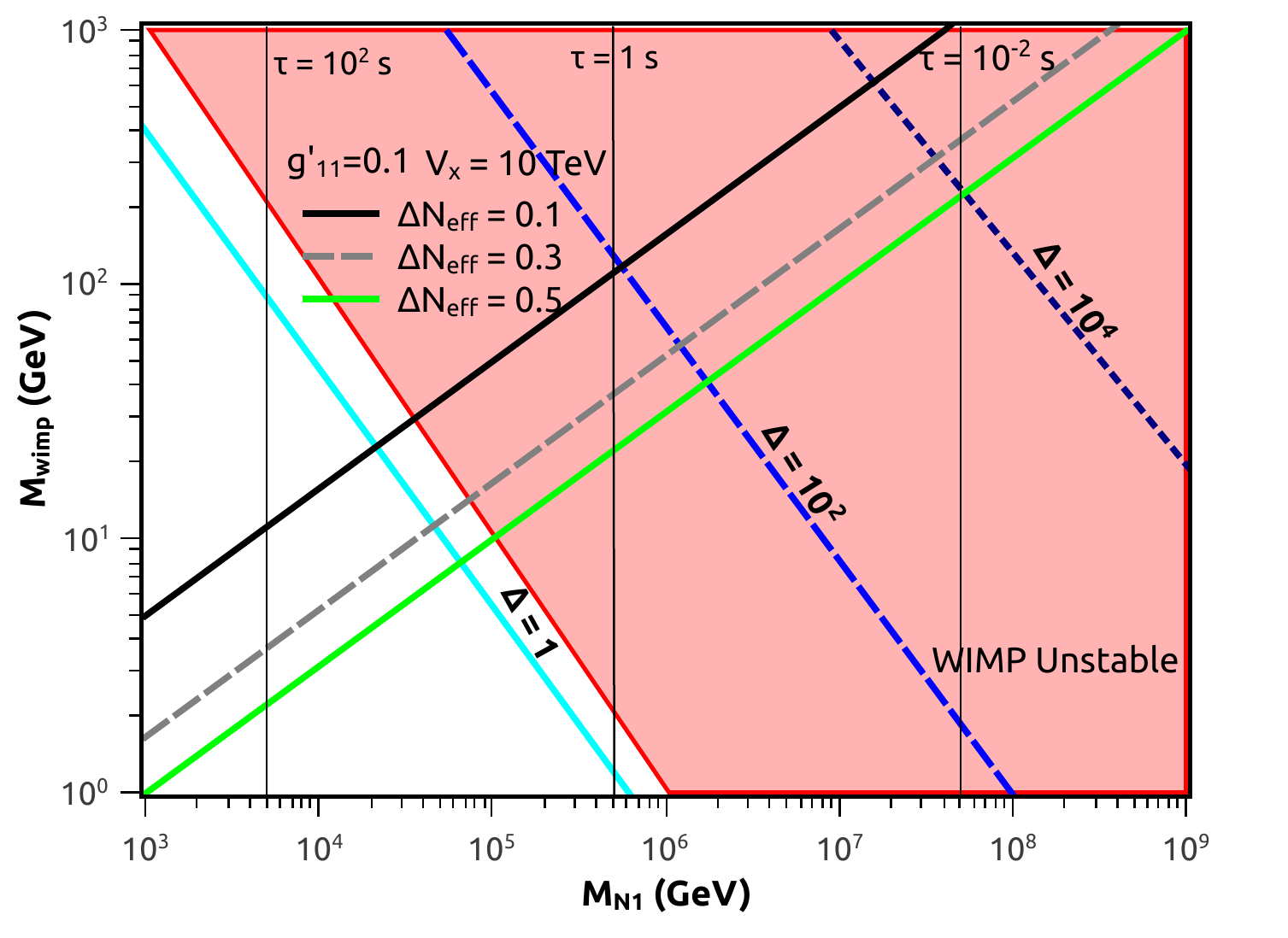}}
\caption{As in Figs.~\ref{g1_large1}-\ref{g1_large2}, but for $g_{11}^\prime=10^{-1}$ with $v_{\chi}^{\prime}=1$~TeV (left panel) and $v_{\chi}^{\prime}=10$~TeV (right panel). The red shaded region induces the WIMP decay.}
\label{g1_small}
\end{figure*}

The parameter space of the theory under consideration can be cast as the choice of the masses for the daughter particle $M_{\rm wimp}$, of the mother particle $M_{N_{1R}}$ and of the coupling constant $g_{11}^{\prime}$, which sets the relevant thermal relic densities. To illustrate the range of viable parameter space where we satisfy all of the constraints outlined above and produce an effective enhancement of the relativistic degrees of freedom $\Delta N_{\rm eff}\sim0.1$, we study the mother-daughter particle mass plane $(M_{N1},M_{\rm wimp})$ for fixed values of the coupling $g_{11}^{\prime}$ (respectively $g_{11}^{\prime}=1$ with $v_{\chi}^{\prime}=1$~TeV in Fig.~\ref{g1_large1}, $g_{11}^\prime=1$ with $v_{\chi}^{\prime}=10$~TeV in Fig.~\ref{g1_large2}, $g_{11}^\prime=10^{-1}$ with $v_{\chi}^{\prime}=1$~TeV and lastly $g_{11}^\prime=10^{-1}$ with $v_{\chi}^{\prime}=10$~TeV in Fig.~\ref{g1_small}.
  
For each $(M_{N_{1R}},M_{\rm wimp})$ pair we enforce that the mass fraction of dark matter produced in a relativistic state from the decays of $N_1$ be exactly $f=0.01$. Given the thermal relic density as calculated within a standard cosmological setup, the abundance of $N_1$ is typically too large to only produce 1\% of the WIMP density. As a result, across most of the parameter space, and especially for small values of $g_{11}^{\prime}$ we postulate that an entropy injection episode occurred between the relatively high temperature at which the $N_1$ froze out and the time of decay (the latter is indicated by vertical lines in the figures). Even though we could have constructed explicit reheating scenarios that could accomplish this, we decided to take a model-independent view, and we phenomenologically parametrize the effect of the entropy injection episode by means of a dilution factor $\Delta$. In other words, the standard thermal relic density $\Omega_{N1}\to\Omega_{N1}/\Delta$ as a result of the larger entropy density. $\Delta=1$ reproduces the standard cosmological model.

A value of $\Delta<1$ indicates that the $N_1$ relic density is too small to provide enough relativistic WIMPs. In this case, one could also postulate cosmologies where the standard thermal relic density is affected and, in particular, enhanced with respect to the standard calculation. Example scenarios include partial non-thermal production for the $N_1$ themselves, or a modified Hubble expansion rate $H\sim T^{2+\alpha} $ with $\alpha>0$ (for example, in the kination-dominated phase of certain quintessence models, $\alpha=1$ \cite{kination} with large potential enhancements of the thermal relic density \cite{profumoullio}). We typically find, however, that across most of the parameter space $\Delta\gg1$.

Figs.~\ref{g1_large1}-\ref{g1_large2} show the mother-daughter parameter space for a relatively large  coupling $g_{11}^\prime=1$. In Fig.~\ref{g1_large1} we used $v_{\chi}^{\prime}=1$~TeV whereas in Fig.~\ref{g1_large2} $v_{\chi}^{\prime}=10$~TeV. The parameter space delimited by the red shaded region in all figures induce the WIMP decay.
For a fixed $g_{11}^\prime$, this decay might be prevented by by increasing the mass of $N_L$, i.e. the value of $v_{\chi}^{\prime}$. For this same reason Fig.~\ref{g1_large2} has a larger parameter space that does induce the decay of the WIMP. In Fig.~\ref{g1_large1} we find a line across the parameter space where all of the constraints are satisfied, and where $\Delta N_{\rm eff}=0.1$ in a standard cosmology for WIMP masses in the range between a few GeV and a few tens of GeV, and for $N_1$ masses between 10 and 100 TeV. Larger $N_1$ masses require increasingly larger entropy suppression factors $\Delta$, and larger WIMP masses to obtain the desired enhancement to $\Delta N_{\rm eff}$. Notice in Fig.~\ref{g1_large2} that when we increase the scale of symmetry breaking larger masses are allowed, but greater entropy suppressions are required though.

In Fig.~\ref{g1_small} we illustrate the situation for smaller values of $g_{11}^\prime=10^{-1}$ with $v_{\chi}^{\prime}=1$~TeV (left panel) and $v_{\chi}^{\prime}=10$~TeV (right panel). From the left panel of Fig.~\ref{g1_small} we can see that this dark radiation scenario is excluded because the parameter space that mimics the number of effective neutrinos induces the WIMP decay. However, increasing the scale of symmetry breaking up to 10TeV a viable region opens up that is able to reproduce the measured value $\Delta N_{eff} ~ 0.1-0.5$ while evading all constraints. If we had used smaller values for $g_{11}^\prime$ instead, entropy suppression factors would have been significantly larger, ranging from $10^6$ all the way up to $10^{16}$, but such smaller couplings are rather disfavored because they induce the WIMP decay.
 
\section{CONCLUSIONS}

It has been proposed recently in the literature that some fraction of the dark matter particles produced non-thermally  could mimic the effect of additional relativistic neutrino species. In Sec. \ref{darkradiation} we examined the conditions which any particle physics model should satisfy in order to offer a plausible way to reproduce $\Delta N_{eff}$ through this non-thermal WIMP setup, namely: ({\it i}) mass of the mother particle should be much larger than the decay products; ({\it ii}) lifetime smaller than $10^4$~s or so, but longer than the epoch of WIMP freeze-out; ({\it iii}) just a small fraction ($\sim 1\%$ or smaller) of the WIMP should be produced with large kinetic energies. In this work we have investigated if this dark radiation scenario with 100~GeV WIMPs is plausible as an electroweak extension of the Standard Model that has $SU(3)_L$ triplets of scalars in its particle content. In our model the mass of the mother particle is determined by a see-saw mechanism with a high-scale Majorana mass term unrelated to the weak-scale, while the mass of the WIMP is set by the scale of symmetry breaking of the model. This means that the huge mass splitting required for this mechanism to work can be easily achieved.  We found that in the model under investigation a very wide range of parameters are capable of producing $\Delta N_{eff}\simeq 0.5$ with a 100~GeV WIMP while still obeying BBN and structure formation bounds.   


\begin{acknowledgements}
The authors thank Lorenzo Ubaldi, Wan-Il Park, Jason Kumar, Kuver Sinha and Rouzbeh Allahverdi for helpful comments and are indebted to Patrick Draper. We also thank CETUP workshop and PPC conference for their hospitality. This work is partly supported by the Department of Energy under contract DE-FG02-04ER41286 (SP), and by the Conselho Nacional de Desenvolvimento Cient\'{\i}fico e Tecnol\'{o}gico (CNPq) ( C.A.S.P, P.S.R.S, F.S.Q ).
\end{acknowledgements}



\begin{thebibliography}{}

\bibitem{Planck} Planck Collaboration,  arXiv:1303.5076.

\bibitem{Cepheid} A. G. Riess {\it  et al.}, Astrophys. J. 730(2011), 119.

\bibitem{Planckanalyses} N. Said, E. Di Valentino, M. Gerbino, arxiv:1304.6217.

\bibitem{Neffevidence} Z. Hou {\it et al.}, arxiv:1212.6267; K. T. Story {\it et al.}, arxiv:1210.7231; M. Archidiacono, E. Giusarma, A. Melchiorri, and O. Mena, arxiv:1303.0143; A. Conley {\it et al.}, Astrophys. J. Suppl. 192, 1 (2001).

\bibitem{susynonthermal} M. Kawasaki and T. Moroi, Prog.Theor.Phys. {bf 93}, 879 (1995),  J. L. Feng, A. Rajaraman, and F. Takayama, Phys. Rev. Lett. {\bf 91}, 011302  (2003),   J. L. Feng, A. Rajaraman, and F. Takayama, Phys. Rev. D {\bf 68} 063504  (2003), J. A. R. Cembranos, J. L. Fend, A. Rajaraman, and F. Takayama, Phys. Rev. Lett. {\bf 95}, 181301 (2005), L. Covi, J. Hasenkamp, S. Pokorski, J. Roberts, JHEP 0911: 003 (2009).

\bibitem{Neffpaper0} W. Fischler and J. Meyer, Phys.Rev. D83 063520 (2011), [arXiv:1011.3501].

\bibitem{Neffpaper} D. Hooper, Farinaldo S. Queiroz, Nickolay Y. Gnedin, Phys. Rev. D{ bf 85}, 063513 (2012).

\bibitem{Neffpaper2} C. Kelso, S. Profumo, Farinaldo S. Queiroz,  Phys.Rev. D 88 (2013) 023511, [arXiv:1304.5243].

\bibitem{Neffpaper3} C. Brust, D. E. Kaplan, and M. T. Walters, arxiv:1303.5379, P. Graf and F. D. Steffen, arxiv:1302.2143, K. J. Bae, H. Baer, A. Lessa, arxiv:1301.7428, U. Franca, R. A. Lineros, J. Palacio, S. Pastor, arXiv:1303.1776,  L. Covi, J. Hasenkamp, S. Pokorski, J. Roberts, JHEP 0911: 003,2009,
P. Graf and F. D. Steffen, arxiv:1302.2143,  W. Fischler and J. Meyers, Phys. Rev. D {\bf 83}, 063520 (2011) ,  J. Hasenkamp, J. Kersten, arXiv:1212.4160.

\bibitem{331higgs} W. Caetano, C. A. de S. Pires, P. S. Rodrigues da Silva, D. Cogollo, Farinaldo S. Queiroz, [arXiv:1305.7246];A. Alves, E. Ramirez Barreto, A. G. Dias, C. A. de S. Pires, F. S. Queiroz, P. S. Rodrigues da Silva ,Phys. Rev. D {\bf 84} 115004  (2011); A.  Alves, A. G. Dias, E. R. Barreto, C. A. de S. Pires, Farinaldo S. Queiroz, P. S. Rodrigues da Silva, Eur. Phys. J. C{\bf 73}, 2288  (2013) 2288.

\bibitem{others}	
F.  Cuypers, S.  Davidson, Eur. Phys. J. C{ \bf 2}, 503 (1998),  A. Alves, E. R.  Barreto, A. G. Dias, Phys. Rev.  D{\bf 84}, 075013  (2011), D. Cogollo, A.Vital de Andrade, F. S. Queiroz, P. Rebello Teles, Eur. Phys. J. C{\bf 72}, 2029 (2012), E. R.  Barreto, Y. A. Coutinho, J. Sa Borges, Phys. Rev. D{ \bf 83}, 054006  (2011),  E. R.  Barreto, Y. A. Coutinho, J. Sa Borges, Phys. Rev. D {\bf 83}, 075001 (2011), M. D. Tonasse,  Phys. Lett. B{\bf 718}, 86 (2012).

\bibitem{331DM_1} C.  A. de S. Pires, and P. S. Rodrigues da Silva, J. Cosmol.Astropart.Phys. {\bf 12}, 012 (2007) .

\bibitem{331DM_2} J. K. Mizukoshi, C. A. de S. Pires, F. S. Queiroz, P. S. Rodrigues da Silva, Phys. Rev. D {\bf 83} , 065024 (2011).

\bibitem{331DM_3} Stefano Profumo and Farinaldo S. Queiroz, [arXiv:1307.7802].

\bibitem{families}P. H.  Frampton,  Phys. Rev. Lett. {\bf 69}, 2889 (1992).


\bibitem{susysneutrino}M. Kawasaki and T. Moroi, Phys. Lett. B {\bf 346}, 27 (1995), J.  L. Feng, S. Su, F. Takayama, Phys.Rev. D {\bf 70}, 063514 (2004),  T. Kanzaki, M. Kawasaki, K. Kohri, and T. Moroi, Phys.Rev.D {\bf 75}, 025011 (2007) .

\bibitem{hadroniclimits} P. Gondolo, G. Gelmini, and S. Sarkar,  Nucl.Phys. B{ \bf 392}, 111 (1993), K. Kohri, Phys. Rev. D {\bf 64}, 043515 (2001) ,  M. Kawasaki, K. Kohri, and T. Moroi, Phys.Rev.D {\bf 71}, 083502 (2005),  M. Kawasaki, K. Kohri, and T. Moroi, Phys. Lett.B {\bf 625}, 7 (2005) 7-12, T. Kanzaki and M. Kawasaki, Phys.Rev.D {\bf 76}, 105017 (2007).

\bibitem{ecq}
C. A. de S, Pires, O. P. Ravinez  Phys. Rev. D {\bf 58}, 035008, Phys. Rev. D{ \bf 60}, 075013 (1999).

\bibitem{Dong} P.V. Dong, T. Phong Nguyen, D.V. Soa, [arXiv:1308.4097]; P.V. Dong, H.T. Hung, T.D. Tham, Phys.Rev. D87 (2013) 115003, [arXiv:1305.0369].

\bibitem{model331}
 M. Singer, J. W. F. Valle and J. Schechter, Phys.
Rev. D {\bf 22}, 738 (1980), J. C. Montero, F. Pisano, and V. Pleitez, Phys. Rev. D {\bf 47},
2918 (1993); R. Foot, H. N. Long, and T. A. Tran, Phys.
Rev. D {\bf 50}, R34 (1994); H. N. Long, ibid. {\bf 54}, 4691 (1996).

\bibitem{minimalversion}
F. Pisano and V. Pleitez, Phys. Rev. D {\bf 46} , 410 (1992); F. Queiroz, C.A. de S.Pires, P.S.Rodrigues da Silva, Phys.Rev. D82 (2010) 065018, [arXiv:1003.1270].

\bibitem{331nu} A. G. Dias, C. A. de S.Pires, P. S. Rodrigues da Silva, Phys. Lett. B{\bf 628}, 85 (2005).

\bibitem{GCpaper} D. Hooper and T. Linden, Phys.Rev. D{\bf 84}, 123005 (2011); Dan Hooper, I. Cholis, T. Linden, J. Siegal-Gaskins, T. Slatyer, arXiv:1305.0830; Dan Hooper, C. Kelso, Farinaldo S. Queiroz, arXiv:1209.3015; Dan Hooper, T. Linden, Phys. Rev. D{ \bf 83}, 083517 (2011),  Dan Hooper, L.  Goodenough, Phys.Lett. B {\bf 697}, 412 (2011) .

\bibitem{GC331} J. D. Ruiz-Alvarez, C. A. de S. Pires, Farinaldo S. Queiroz, D. Restrepo, P. S. Rodrigues da Silva, Phys. Rev. D{\bf 86}, 075011 (2012) .

\bibitem{micromegas} G. Belanger, F. Boudjema, A. Pukhov, and A. Semenov, Comput. Phys. Commun. 176, 367 (2007); G. Belanger, F.Boudjema, A. Pukhov, and A. Semenov, Comput. Phys. Commun. 180, 747 (2009); G. Beelanger, F. Boudjema, A. Pukhov, and A. Semenov, [arXiv:1005.4133].

\bibitem{kination} P.~G.~Ferreira and M.~Joyce,
  Phys.\ Rev.\ D {\bf 58}, 023503 (1998)
  [astro-ph/9711102].

\bibitem{profumoullio} 
  S.~Profumo and P.~Ullio,
  JCAP {\bf 0311}, 006 (2003)
  [hep-ph/0309220].




\end{thebibliography}
\end{document}